# Formation and Dynamical Evolution of the Neptune Trojans – the Influence of the Initial Solar System Architecture

P. S. Lykawka,[1*] J. Horner,[2] B. W. Jones[3] and T. Mukai[4]

[1] International Center for Human Sciences (Planetary Sciences), Kinki University, 3-4-1 Kowakae, Higashiosaka, Osaka, 577-8502, Japan

[2] Dept. of Physics, Science Laboratories, University of Durham, South Road, Durham, DH1 3LE, United Kingdom

[3] Dept. of Physics and Astronomy, The Open University, Walton Hall, Milton Keynes, MK7 6AA, United Kingdom

[4] Kobe University, 1-1 rokkodai-cho, nada-ku, Kobe 657-8501, Japan

---

[*] E-mail address: patryksan@gmail.com
Previous address: Dept. of Earth and Planetary Sciences, Kobe University, 1-1 rokkodai-cho, nada-ku, Kobe 657-8501, Japan.

**ABSTRACT**

Current models of Solar system formation suggest that the four giant planets accreted as a significantly more compact system than we observe today. In this work, we investigate the dynamical stability of pre-formed Neptune Trojans under the gravitational influence of the four giant planets in compact planetary architectures, over 10 Myr. In our modelling, the initial orbital locations of Uranus and Neptune ($a_N$) were varied to produce systems in which those planets moved on non-resonant orbits, or in which they lay in their mutual 1:2, 2:3 and 3:4 mean-motion resonances (MMRs). In total, 420 simulations were carried out, examining 42 different architectures, with a total of 840000 particles across all runs. In the non-resonant cases, the Trojans suffered only moderate levels of dynamical erosion, with the most compact systems (those with $a_N \leq 18$ AU) losing around 50% of their Trojans by the end of the integrations. In the 2:3 and 3:4 MMR scenarios, however, dynamical erosion was much higher with depletion rates typically greater than 66% and total depletion in the most compact systems. The 1:2 resonant scenarios featured disruption on levels intermediate between the non-resonant cases and other resonant scenarios, with depletion rates of the order of tens of percent. Overall, the great majority of plausible pre-migration planetary architectures resulted in severe levels of depletion of the Neptunian Trojan clouds. In particular, if Uranus and Neptune formed near their mutual 2:3 or 3:4 MMR and at heliocentric distances within 18 AU (as favoured by recent studies), we found that the great majority of pre-formed Trojans would have been lost prior to Neptune's migration. This strengthens the case for the great bulk of the current Neptunian Trojan population having been captured during that migration.

# 1 INTRODUCTION

The population of objects collectively known as Neptunian Trojans represent the most recent addition to the menagerie of objects that make up our Solar system. These objects orbit the Sun within Neptune's 1:1 mean-motion resonance (MMR), with a characteristic motion where they drift in a periodic manner around the so-called L4 and L5 Lagrange points, which are located 60° ahead and behind the planet in its orbit (Murray & Dermott 1999). Such periodic motion is known as *libration*. Though the existence of such objects has long been postulated, the first was not identified until 2001 (Chiang et al. 2003). That object, 2001 QR322, surely represents the first of a huge population of objects trapped within Neptune's 1:1 MMR (Sheppard & Trujillo 2006). Indeed, since its discovery, a further five such objects have been found, all librating with 2001 QR322 around Neptune's leading L4 Lagrange point (Zhou, Dvorak & Sun 2009; Lykawka et al. 2009).

As described elsewhere (Nesvorny & Vokrouhlicky 2009; Lykawka & Horner 2010; Lykawka et al. 2009), the Neptunian Trojans represent a unique window into the formation of our Solar system. The distribution of these objects through eccentricity and inclination space has already thrown up significant surprises (such as the apparent excess of highly inclined Trojans, $i$>5°, over what was previously expected), and can be directly linked to the behaviour of the giant planets during the final stages of their evolution. In addition, we have also found that the Neptune Trojans could represent a significant source of objects wandering on dynamically unstable orbits in the outer Solar system, known as the Centaurs, which in turn are widely accepted to be the primary source of short-period comets (for more information, see Horner, Evans & Bailey 2004; Horner & Lykawka 2010).

In our earlier work (Lykawka et al. 2009, hereafter Paper I), we examined the evolution of populations of objects that either formed within Neptune's Trojan clouds, or were captured to them, as the planet followed one of four different migration scenarios through the outer Solar system. Here, we turn our attention to the various possible architectures that might have existed shortly before the onset of such migration – at an imaginary $t_0$ for that migration. For a wide variety of such architectures, we examine the structure and stability of Neptune's Trojan clouds, in order to determine what effect, if any, the initial architecture of the planetary system would have had on the surviving number and distribution of in-situ Neptunian Trojans at the end of Neptune's assembly, before its outward migration.

In Paper I, we obtained results that showed that the initial positions of the giant planets, relative to one another, play an important role in determining the stability of primordial (pre-formed) Neptune Trojans. In particular, one of the scenarios considered in that work had Uranus and Neptune located initially close to their mutual 3:4 MMR, with the planets proceeding to cross a number of other strong higher-order mutual MMRs[1] as they migrated outward (Lykawka & Horner 2010). In that scenario, over 99% of pre-formed Trojans were shed within the first 3-5 Myr of the planets 50 Myr migration. A closer analysis of the survival fractions for the pre-formed Trojans in that work suggests that, the closer Uranus and Neptune are to one another at the beginning of their evolution, the more likely it is that a significant fraction of the pre-formed Trojan clouds will be destabilised during the migration of those planets. This result supports that obtained in pioneering work by Gomes (1998), who noted that the stability of Neptunian Trojans was strongly influenced by the mutual separation of Uranus and Neptune. However, to our knowledge, other than this pioneering work, the influence of the initial architecture of the outer Solar system on the survival of primordial Neptunian Trojan population has never been examined.

---

[1] The strongest MMRs are usually those with the smallest resonance order, which is obtained from |p-q| for a MMR defined as p:q (e.g., 1:2, 2:3, 3:4, etc. are examples of 1st order MMRs). For a more complete and detailed discussion on resonance strength and its dependence on orbital elements and resonance order, please refer to Gallardo (2006) and Lykawka & Mukai (2007).

It is well known that, at the current day, the L4 and L5 points in the orbits of Saturn and Uranus are do not offer sufficient dynamical stability to house a substantial population of stable Trojans, while those of Jupiter and Neptune may house more objects than orbit within the main asteroid belt (Sheppard & Trujillo 2006). Assuming that a significant fraction of the Neptunian Trojans were transported with the planet, having formed with it, rather than being captured en-route, the initial architecture of the system must have been such that the Neptunian clouds were stable before migration began. However, a great number of feasible initial architectures for the outer Solar system could lead to great instability in the Neptunian Trojans. Therefore, a study of the stability of these regions as a function of the initial, pre-migration, orbits of the planets can be enlightening as to which architectures could lead to a Trojan system compatible with that observed today. In addition, by investigating the various plausible initial planetary architectures, we can also place constraints on how the local environment for the formation and early evolution of Neptune Trojans depends on the compactness of the system. Finally, since a number of recent studies of the early orbital evolution of giant planets embedded in gaseous disks have reported that such planets could evolve in mutual MMRs, and even remain in them after the dissipation of the disk gas (Thommes 2005; Papaloizou & Szuszkiewicz 2005; Morbidelli et al. 2007), it is clearly important to consider the stability of Neptune Trojans in situations where Uranus and Neptune lie on mutually resonant orbits. As such, in this work, we examine a variety of planetary architectures that match those that could have arisen after the removal of gas from the proto-planetary disk (our time $t = 0$), studying a variety of non-resonant, resonant, and highly compact systems. This allows us to determine the effect the various scenarios would have on the overall stability, and dynamical excitation, of pre-formed Neptune Trojans.

In Section 2, we detail the techniques used to study the behaviour of in-situ Trojans under a variety of different planetary architectures. In Section 3, we give the results of our simulations, which are discussed in detail in Section 4. Finally, in Section 5, we present our key conclusions, and discuss our plans for future work on this topic.

## 2 MODELLING

A great number of models have been proposed to explain the formation of the outer planets (Kokubo & Ida 2002; Goldreich, Lithwick & Sari 2004; Alibert et al. 2005). In the last decade, these have incorporated the concept that the planets might have migrated over significant distances before reaching their current locations (Fernandez & Ip 1984; Hahn & Malhotra 1999; Levison et al. 2007). Such migration is invoked in order to explain a number of features of the Solar system, including the orbital distribution of objects captured in Neptune's MMRs, and the dynamical depletion vast swathes of the asteroid belt (Liou & Malhotra 1997; Minton & Malhotra 2009). The number of permissible pre-migration architectures allowed by these models is almost infinite, though it is thought that Neptune must have migrated outwards over a significant number of astronomical units distance in order to produce the highly excited resonant populations we see today (Hahn & Malhotra 2005; Lykawka & Mukai 2008 and references therein).

In this work, we consider 42 separate initial architectures for the outer Solar system, and pay particular attention to the relationship between the initial orbits of Uranus and Neptune and the survivability of objects formed around Neptune's L4 and L5 Lagrange points. In each case, one thousand massless Trojan particles were placed at L4, and a further thousand were spread around L5. These particles were distributed randomly within the clouds such that their semi-major axes lay within 0.001 AU of that of Neptune, and their eccentricities were less than 0.01. For each particle, the inclination was set, in radians, to equal one half of its orbital eccentricity. Finally, they were distributed so that their maximum initial libration would take them no further than 10° from their Lagrange point.

In every case, the particles were followed for a period of 10 Myr under the influence of the four giant planets, Jupiter, Saturn, Uranus and Neptune, using the *Hybrid* integrator within the dynamics package MERCURY (Chambers 1999). Though 10 Myr is only a small fraction of the age of the Solar system, and so might seem a somewhat arbitrary choice of integration length, it is more than sufficient to allow us to draw important conclusions on the evolution of the planetary Trojan populations over the course of their migration. Particles were removed from the simulation when they collided with a massive body, or passed beyond 40 AU from the Sun. Once the integrations were complete, the orbital elements for each particle were examined using the software package RESTICK (Lykawka & Mukai 2007) to determine the Trojan lifetime[2] duration and their resonant properties.

In this work, we primarily examine the case where Uranus and Neptune are located in a non-resonant configuration, with a reasonable mutual separation. In addition, we consider three additional scenarios in which the two planets start with an initial separation that places them in particular mutual MMRs. The details of the various architectures examined are as follows:

**Architecture A:** Uranus and Neptune form in a non-resonant configuration, separated typically by 7.5 times their mutual Hill radius.

Taking inspiration from the findings of previous studies of planet formation, the mutual locations of Neptune and Uranus are determined using the following equation:

$$a_N = a_U + f R_{mH} \qquad \text{Eq. 1,}$$

where $a_N$ and $a_U$ are the semimajor axis of Neptune and Uranus (in AU), and $R_{mH}$ is the mutual Hill radius of both planets (determined using Eq. 2). $f$ is a simple multiplicative term used to determine the number of Hill radii separating the planets. The value of $R_{mH}$ is defined by

$$R_{mH} = \left(\frac{a_N + a_U}{2}\right)\left(\frac{m_N + m_U}{3M_*}\right)^{\frac{1}{3}} \qquad \text{Eq. 2,}$$

where $m_N$ and $m_U$ are the current masses of Neptune and Uranus, measured in terms of the Solar mass (natural units, such that $M_* = 1$). From the combination of Equations 1 and 2, it is can easily be shown that

$$a_U = a_N \frac{2 - fN}{2 + fN} \qquad \text{Eq. 3, where } N = \left(\frac{m_U + m_N}{3}\right)^{\frac{1}{3}} \sim 0.031657.$$

So, given an initial location for Neptune, it is trivial to calculate where Uranus would reside. Typical planet formation models assume that the initial separation between Uranus and Neptune is somewhere between 5 and 10 mutual Hill Radii (Ida & Lin 2004 and references therein), and so, for runs involving such architecture, we take the intermediate value $f = 7.5$ to determine our standard mutual separation.

---

[2] An object is defined as occupying a Trojan orbit if its resonant angle (the angular separation of the object from Neptune around the orbit of that planet, measured in degrees) undergoes libration, rather than circulation (in other words, if the resonant angle oscillates around a specific value (60º, for example), rather than merely circulating around from 0 to 360º). Such librational behaviour results in the object remaining a significant angular separation from Neptune during its orbital motion, and therefore prevents the object being gravitationally scattered by the planet.

13 different values of $a_N$ were tested, ranging from 15 to 27 AU in 1 AU increments. Uranus was moved outward in lockstep with the location of Neptune, so that the two planets were 7.5, 9, 10, and 11 $R_{mH}$ apart for $a_N$ = 15-24 AU, 25 AU, 26 AU and 27 AU, respectively. For these final three values of $a_N$, slightly larger values of $f$ were necessary in order to meet the requirement that the initial heliocentric distance of Uranus be less than its current value, to remain compatible with models that require that planet to migrate outward, rather than inward.

**Architecture B:** Uranus was placed so that it lay within the 1:2 MMR with Neptune. Here, 9 variants in $a_N$ were used, ranging from 19 to 27 AU (values of $a_N$ of 18 AU or less would have placed Uranus too close to Saturn to fit with current planet formation models).

**Architecture C:** Uranus was placed so that it lay within the 2:3 MMR with Neptune. In this case, 11 variants of $a_N$ were used, ranging from 15 to 25 AU (values of $a_N$ of 26 or 27 AU were excluded since they would place Uranus further from the Sun than its current location (19.2 AU), and so would not fit with current models of planet formation and migration).

**Architecture D:** Uranus was located so that it lay within the 3:4 MMR with Neptune. Here, 9 variants of $a_N$ were used, ranging from 15 to 23 AU (as with type C, if values of $a_N$ of 24 AU or greater were used, this would place $a_U$ beyond Uranus' current semimajor axis).

In all runs of all architectures, Jupiter and Saturn were initially placed at semi-major axes of 5.4 and 8.7 AU, respectively, values chosen to fit their supposed pre-migration locations. Each set-up was tested ten separate times, with the planets being initialised at random locations in their orbit, in order to provide a fair test, leading to 420 simulations being carried out, using 840,000 particles in total. The planets did not migrate during the course of the simulations, but were wholly gravitationally interacting.

**3 RESULTS**

As an initial test of the data we obtained, we first analysed the orbital evolution of the giant planets over the 10 Myr of each of the simulations described in Section 2. In the non-resonant integrations, periodic small-scale orbital variations were observed, whilst in the resonant scenarios the orbits of Uranus and Neptune often became slightly excited. In the great majority (395) of these cases, the giant planets experienced negligible variation in orbital elements. The data obtained in these 395 "stable" runs provide the main thrust of this work, and will be discussed in Section 3.1, below. A small minority (25) of the initial runs (typically those involving the most compact planetary systems) resulted in perturbations that caused at least one of the giant planets to undergo chaotic dynamical evolution (characterised by large excursions in semimajor axis, or large variations in orbital eccentricity). The data obtained from these 25 chaotic integrations was rejected from the main analysis work, and is considered separately, for completeness, in Section 3.2.

We examined the orbital data from these integrations using the RESTICK analysis tool (Lykawka & Mukai 2007). This allowed us to identify Trojans by checking the libration of the resonant angle with respect to Neptune ($\phi = \lambda - \lambda_N$, where $\lambda$ and $\lambda_N$ are the mean longitudes of the object and Neptune, respectively). We then obtained the distribution of Trojan objects in both element space ($a$-$e$-$i$) and as a function of their resonant properties: namely the location of the centre of libration, the libration period, and the libration amplitude[3]. The detection method was calibrated such that Trojans moving both on tadpole (librating around the L4 and L5 Lagrange points) and horseshoe orbits could be found, so long as their dynamical Trojan lifetime was of the order ~0.2 Myr. We also recorded both the number of Trojan objects controlled by each giant planet, and their particular resonant properties, at the end of the 10 Myr integrations for each of the 420 runs.

---

[3] The libration amplitude refers to the maximum angular displacement from the centre of libration achieved during the object's resonant motion (i.e. is equal to half the full extent of its angular motion).

After grouping the results of individual runs according to type of planetary architecture involved (see Section 2), we determined the survival fractions of Trojan objects as a function of Neptune's initial location at the end of the 10 Myr integration. We also calculated the approximate number of Trojans as a function of time at 0.1 Myr intervals over the total integration time, 10 Myr, using a more primitive algorithm unrelated to RESTICK. In these particular calculations, objects moving on orbits with semimajor axes within ±0.25 AU of a given giant planet, and simultaneously incapable of undergoing close encounters with any other giant, were considered to be Trojans of that planet. Objects were checked for such behaviour at each data output step (2 kyr intervals). The number of Trojans was then calculated at 0.1 Myr intervals by summing the number of individual objects that satisfied the above orbital condition over 50 consecutive time steps. The total number of Trojans identified this way was checked, and found to be in good agreement with that obtained by RESTICK at distinct times over 10 Myr. We stress that the number of Trojans obtained as a function of time was intended to yield only statistical results on these objects for each of the architectures considered in this work (e.g., decay lifetimes), which we discuss in detail below.

Practically all those Trojans that survived through the integrations retained low eccentricities and inclinations ($e \sim i < 0.02$), thus preserving their initial cold dynamical nature. However, in the most compact planetary systems, a significant and increasing fraction of Trojans acquired eccentricities in the range $e < 0.05$-$0.07$ and $e < 0.15$-$0.2$ for non-resonant and resonant systems, respectively. This feature can be clearly seen in results in Table 1, when one examines the variation in the average eccentricity values as a function of $a_{N0}$. However, in contrast to this moderate excitation in eccentricity, no appreciable excitation in inclination was observed among all the studied Trojan populations.

Finally, those Trojans that survived for 10 Myr showed libration amplitudes ranging from very few to about 30-35°, irrespective of the initial planetary architecture. It seems reasonable to conclude, therefore, that the great bulk of the pre-formed (rather than captured) Neptune Trojan population that survived until the onset of planetary migration would have librated about the Lagrange points with maximum angular displacements no greater than 70°. On longer timescales, this maximum libration value might slightly decrease as a result of the ejection of unstable objects with Trojan lifetimes greater than 10 Myr.

### 3.1 QUASI-STABLE SYSTEMS
The survival fraction of pre-formed Trojans as a function of Neptune's initial location at the end of 10 Myr are illustrated in Figure 1 for the non-resonant and the three resonant configurations. These fractions represent the number of Neptune Trojans remaining, prior to the onset of planetary migration, for various planetary configurations during the early Solar system.

For non-resonant configurations (Architecture A), the number of Neptune Trojans that survive for the full duration of our integrations is directly related to both the initial semi-major axis of Neptune and its initial mutual distance from Uranus (black curve in Fig. 1). The wider the initial separation, and the greater the semi-major axis of Neptune, the more objects survive. The fraction retained varies from approximately 45%, in the case of the most compact system ($a_{N0} = 15$ AU), to >95% in wider systems ($a_{N0} > 21$ AU). Interestingly, as a result of our data being obtained from ten distinct runs for each specific system investigated, the small dip at $a_{N0} = 18$ AU seems to be a real feature (rather than the result of statistical noise). Although it is plausible that resonant interactions between Uranus and Neptune could partially destabilise the Trojan population, creating such a feature, in this particular case, those planets were not close to any of their mutual MMRs. However, this scenario represents the only configuration considered in which the initial positions of Saturn and Uranus lie close to their mutual 1:2 MMR while Uranus and Neptune are mutually non-resonant. It is therefore reasonable to interpret the lower survival fraction in the case, $a_{N0} = 18$ AU, as being directly related to the resonant interactions between Saturn and Uranus, which typically affected the semimajor axis and eccentricity of Uranus' orbit, causing in turn a similar, albeit less significant, excitation of Neptune's orbit.

The estimated number of Neptune Trojans as a function of time is illustrated in Figure 2. As we saw in the discussion of survival fractions above, the lifetimes of Neptune Trojans tend to be smaller for systems where Neptune started closer to the Sun, which are also the systems in which Uranus and Neptune are less widely separated. In particular, it can be seen that objects start to leave the Trojan clouds within the first million years of their evolution. This is particularly true in the $a_{N0}$ = 18 AU case, where the near-resonance between Uranus and Saturn adds an extra destabilising influence to the outer Solar system. Particularly in the case of the more compact planetary configurations, with $a_{N0}$ = 15-18 AU, the decay curves suggest that the population of Neptune Trojans will continue to decrease as a function of time beyond the end of the integrations at 10 Myr. This would clearly lead to a substantial dynamical erosion of the population in these less stable scenarios, such that the longer the wait before significant planetary migration, the fewer native Trojans would remain under Neptune's control.

The dynamical evolution of Neptune Trojans for the three resonant configurations (Architectures B-D) yielded results remarkably distinct from the non-resonant ones discussed before. In these resonant cases, we find that in general the survival fraction of Trojans is much smaller than in the non-resonant scenarios. Similarly, the number of Neptune Trojans decreases much faster for the resonant architectures during the integrations (Fig. 2). This is particularly true for the 2:3 and 3:4 MMR configurations (the green and red curves in Fig. 1), for which the decay of the Trojan population was especially rapid. In these cases, the Trojan population remaining after 10 Myr was, *at best*, just 35% of the initial number, and in some cases, the entire population of 2000 particles was shed well before the end of the simulation. The survival fraction in these cases appears to be correlated with Neptune's initial semimajor axis. The greater the distance between Neptune and the Sun (and, given that the planets were near-resonant, the greater the separation of Uranus and Neptune), the greater the survival of the Neptune Trojans. This behaviour can itself be divided into two regimes. When the initial semimajor axis of Neptune is less than 18 AU, all integrations yield survival fractions below 15%. However, when the planet is beyond that distance, the survival fractions range between 10% and 35%. Therefore, it is possible that the great majority (or even the entire) population of Trojans formed locally with Neptune was lost if the planet evolved to an orbit within ~18 AU of the Sun, and lay close the 2:3 or the 3:4 MMR with Uranus, during the early Solar system.

Simulations in which Uranus and Neptune lay close to their mutual 1:2 MMR yielded survival fractions that show no clear trend as a function of the initial heliocentric distance of Neptune (the blue curve in Fig. 1), but instead fall between the extremes of the non-resonant scenarios, on the one hand, and the 2:3 and 3:4 MMR scenarios on the other. This behaviour was also noted in the decay curves (Fig. 2) for Trojans in those systems. This suggests the effect of the 1:2 MMR was in general not so disruptive for the Neptune Trojans when compared to the behaviour seen for the 2:3 and 3:4 MMR configurations. The one exception to this result can be seen in both Figs 1 and 2 at around $a_{N0}$ = 22 AU. When the planets are set up such that Neptune is placed at that distance, and Uranus is located in the 1:2 MMR, an abrupt drop in the number of Trojan survivors occurs. We earlier discussed a similar reduction in survival fraction that was observed in the non-resonant case, when Neptune was located at $a_{N0}$ = 18 AU. Once again, it turns out that when Neptune is at 22 AU and Uranus placed near the 1:2 MMR, Uranus just happens to lie particularly close to its mutual 1:2 MMR with the planet Saturn. In this case, this resulted in the two 1:2 MMRs driving the excitation of the eccentricities of Uranus and Neptune's orbits, and causing them to undergo small, erratic, displacements in semimajor axis. These interactions, combined with the overlapping of resonances in the vicinity of the Trojan populations, resulted in a devastating reduction of the likelihood of any of these bodies surviving until the end of the integrations.

A comparison of the retention fraction against Neptunian distance for the three resonant configurations, as seen in Figure 1, suggests that all three scenarios feature retention rates which appear to gradually rise as a function of Neptune's heliocentric distance. The general upward trend is, in each case, marred by a number of abrupt dips at locations dictated by mutual MMRs between Saturn and Uranus. Indeed, substantial drops in the survival curves always occur when Saturn is close to its 1:2 MMR with Uranus, a feature that happens at $a_{N0} \sim 17$, 18, and 22 AU for the 3:4, 2:3 and 1:2 MMR configurations discussed here. This drop is not hugely apparent for the 2:3 and 3:4 MMR cases, since the retention fractions of Trojans in the simulations at adjacent $a_{N0}$ are already particularly low. However, it is clear that in both cases the entire Trojan population is lost prior to the completion of 10 Myr of evolution at these locations. For comparison, the typical retention fraction of Trojans in the 1:2 resonant scenarios is high (mostly over 50%), except when the Saturn-Uranus 1:2 MMR plays a role (at $a_{N0} = 22$ AU), at which point the retention rate falls to just ~5%. Following the curves, we also note further moderate reductions in retention fraction visible at ~ 20, 22, and 27 AU for the 3:4, 2:3 and 1:2 MMR configurations respectively. In each case, the feature coincides with a scenario in which Uranus lies close to its 2:5 MMR with Saturn. From these results, it is clear that versions of our Solar system in which Uranus evolves in close proximity to strong MMRs with Saturn (e.g. 3:4, 2:3, 1:3, etc.) are likely to undergo enhanced depletion in the primordial Neptunian Trojan clouds as a result of both resonance overlapping (which typically leads to unstable chaotic behaviour) and the orbital excitation of Uranus and Neptune themselves.

Taken as a whole, our results clearly suggest that significant or total loss of Neptune Trojans is expected for compact planetary configurations ($a_{N0} \leq 18$ AU), particularly in those systems where the giant planets are located close to mutual MMRs, even within a relatively short time span of 10 Myr. On the other hand, a significant number of stable Trojans could be expected to survive for at least 10 Myr prior to the onset of planetary migration for wider planetary configurations ($a_{N0} > 18$ AU). However, it is clear that Trojan stability is still greatly compromised even in these more widely distributed systems, in those cases where mutual planetary MMRs are invoked. Full details of all runs carried out are given in Table 1, with a wide variety of statistical information on the results obtained in each case.

| $a_{N0}$ (AU) | $a_{U0}$ (AU) | $N_{L4}$ | $N_{L5}$ | $N_H$ | $f_{L4}$ (%) | $f_{L5}$ (%) | $F_H$ (%) | $<e_{L4}>$ | $<e_{L5}>$ | $<A_{L4}>$ (°) | $<A_{L5}>$ (°) |
|---|---|---|---|---|---|---|---|---|---|---|---|
| | | | | | **Non-MMR** | | | | | | |
| 15 | 11.82 | 4226 | 4132 | 685 | 42.3 | 41.3 | 7.6 | 0.025 | 0.028 | 20 | 24 |
| 16 | 12.60 | 4456 | 5029 | 140 | 44.6 | 50.3 | 1.5 | 0.027 | 0.026 | 17 | 17 |
| 17 | 13.39 | 6368 | 5790 | 92 | 63.7 | 57.9 | 0.8 | 0.023 | 0.022 | 22 | 18 |
| 18 | 14.18 | 4561 | 4447 | 116 | 45.6 | 44.5 | 1.3 | 0.019 | 0.018 | 13 | 15 |
| 19 | 14.97 | 8463 | 8370 | 112 | 84.6 | 83.7 | 0.7 | 0.019 | 0.017 | 15 | 17 |
| 20 | 15.76 | 8185 | 8909 | 59 | 81.9 | 89.1 | 0.3 | 0.018 | 0.014 | 15 | 20 |
| 21 | 16.54 | 9132 | 9621 | 35 | 91.3 | 96.2 | 0.2 | 0.011 | 0.010 | 18 | 15 |
| 22 | 17.33 | 10000 | 10000 | 0 | 100.0 | 100.0 | 0.0 | 0.011 | 0.013 | 16 | 18 |
| 23 | 18.12 | 9219 | 9232 | 403 | 92.2 | 92.3 | 2.1 | 0.013 | 0.016 | 18 | 18 |
| 24 | 18.91 | 9602 | 10011 | 46 | 96.0 | 100.1 | 0.2 | 0.011 | 0.008 | 17 | 17 |
| 25 | 18.77 | 10000 | 10000 | 0 | 100.0 | 100.0 | 0.0 | 0.009 | 0.010 | 17 | 14 |
| 26 | 18.89 | 10000 | 10000 | 0 | 100.0 | 100.0 | 0.0 | 0.013 | 0.012 | 17 | 18 |
| 27 | 18.99 | 10000 | 10000 | 0 | 100.0 | 100.0 | 0.0 | 0.012 | 0.013 | 16 | 20 |
| | | | | | **1:2 MMR** | | | | | | |
| 19 | 11.97 | 3320 | 3361 | 642 | 41.5 | 42.0 | 8.8 | 0.021 | 0.019 | 22 | 16 |
| 20 | 12.60 | 7057 | 7203 | 655 | 70.6 | 72.0 | 4.4 | 0.021 | 0.023 | 21 | 21 |
| 21 | 13.23 | 9380 | 9361 | 496 | 93.8 | 93.6 | 2.6 | 0.012 | 0.013 | 17 | 19 |
| 22 | 13.86 | 393 | 322 | 429 | 3.9 | 3.2 | 37.5 | 0.065 | 0.077 | 41 | 52 |
| 23 | 14.49 | 4161 | 4243 | 2044 | 41.6 | 42.4 | 19.6 | 0.028 | 0.028 | 23 | 23 |
| 24 | 15.12 | 6230 | 6250 | 568 | 62.3 | 62.5 | 4.4 | 0.015 | 0.013 | 21 | 22 |
| 25 | 15.75 | 8496 | 8336 | 728 | 85.0 | 83.4 | 4.1 | 0.016 | 0.013 | 17 | 21 |
| 26 | 16.38 | 9916 | 9970 | 49 | 99.2 | 99.7 | 0.2 | 0.010 | 0.011 | 20 | 19 |
| 27 | 17.01 | 8429 | 8668 | 1007 | 84.3 | 86.7 | 5.6 | 0.013 | 0.014 | 21 | 25 |
| **2:3 MMR** | | | | | | | | | | | |
| 15 | 11.45 | 0 | 0 | 0 | 0.0 | 0.0 | - | - | - | - | - |
| 16 | 12.21 | 0 | 0 | 1 | 0.0 | 0.0 | 100.0 | - | - | - | - |
| 17 | 12.97 | 1720 | 1002 | 46 | 17.2 | 10.0 | 1.7 | 0.013 | 0.010 | 13 | 15 |
| 18 | 13.74 | 0 | 0 | 4 | 0.0 | 0.0 | 100.0 | - | - | - | - |
| 19 | 14.50 | 776 | 1350 | 189 | 7.8 | 13.5 | 8.2 | 0.025 | 0.026 | 23 | 12 |
| 20 | 15.26 | 1792 | 1591 | 500 | 17.9 | 15.9 | 12.9 | 0.014 | 0.019 | 23 | 20 |
| 21 | 16.03 | 2508 | 3501 | 1190 | 25.1 | 35.0 | 16.5 | 0.021 | 0.021 | 23 | 22 |
| 22 | 16.79 | 1641 | 1949 | 158 | 16.4 | 19.5 | 4.2 | 0.014 | 0.019 | 15 | 16 |
| 23 | 17.55 | 2721 | 2979 | 285 | 27.2 | 29.8 | 4.8 | 0.019 | 0.017 | 20 | 18 |
| 24 | 18.32 | 3580 | 2705 | 262 | 35.8 | 27.1 | 4.0 | 0.017 | 0.019 | 20 | 23 |
| 25 | 19.08 | 2841 | 1917 | 39 | 28.4 | 19.2 | 0.8 | 0.012 | 0.011 | 14 | 13 |
| **3:4 MMR** | | | | | | | | | | | |
| 15 | 12.38 | 0 | 0 | 0 | 0.0 | 0.0 | - | - | - | - | - |
| 16 | 13.21 | 582 | 1005 | 12 | 5.8 | 10.1 | 0.8 | 0.008 | 0.008 | 6 | 7 |
| 17 | 14.03 | 0 | 0 | 0 | 0.0 | 0.0 | - | - | - | - | - |
| 18 | 14.86 | 6 | 1 | 3 | 0.1 | 0.0 | 30.0 | 0.029 | 0.007 | 29 | 22 |
| 19 | 15.68 | 2741 | 3224 | 51 | 27.4 | 32.2 | 0.8 | 0.010 | 0.010 | 14 | 19 |
| 20 | 16.51 | 939 | 1768 | 67 | 9.4 | 17.7 | 2.4 | 0.015 | 0.013 | 16 | 22 |
| 21 | 17.34 | 1790 | 1752 | 139 | 17.9 | 17.5 | 3.8 | 0.014 | 0.020 | 27 | 25 |
| 22 | 18.16 | 3004 | 2376 | 76 | 30.0 | 23.8 | 1.4 | 0.011 | 0.015 | 17 | 21 |
| 23 | 18.99 | 2005 | 1938 | 13 | 20.1 | 19.4 | 0.3 | 0.010 | 0.012 | 11 | 18 |

**Table 1:** Details of the Neptune Trojans that survived 10 Myr of integration across a variety of planetary architectures (split into four blocks). The initial heliocentric distances of Uranus ($a_{U0}$) and Neptune ($a_{N0}$) were varied, giving a range of non-resonant (Non-MMR) and resonant architectures. Three distinct resonant configurations were studied, with Uranus and Neptune lying close to their mutual 1:2, 2:3, and 3:4 MMRs. $N_{L4}$, $N_{L5}$ and $N_H$ give the number of particles moving on tadpole and horseshoe orbits at the end of the simulations. $f_{L4}$ and $f_{L5}$ also detail the number of tadpole Trojans at 10 Myr, this time expressed as a percentage of the initial population of objects at that location. $F_H$ gives the fraction of the final population of Trojans that are moving on horseshoe orbits in the Trojan cloud at the conclusion of our simulations. The quantities detailed on the right hand, <e> and <A>, give the averaged eccentricity and libration amplitude for the final Trojan populations around each of the Lagrange points (as indicated by the L4 and L5 subscript symbols). Finally, it should be noted that a number of Trojans found around the Lagrangian point L4 (L5) at 10 Myr originally started librating about at the L5 (L4) point. The fraction of objects that experienced such L4-L5 (L5-L4) migrations typically contributed <1% and ~1-5% of the final population found at each Lagrangian point for the most stable and less stable systems, respectively.

## 3.2 CHAOTIC SYSTEMS

In 25 of the 420 integrations carried out, at least one of the giant planets evolved chaotically, as previously mentioned in Section 3. These 25 cases all belonged to the resonant scenarios studied, distributed between the 1:2 (2 runs), 2:3 (14 runs) and 3:4 (9 runs) scenarios. The planetary instability was triggered by mutual 1:2 MMR crossings experienced by Jupiter and Saturn in the majority of these systems. Such crossings typically occurred as a result of non-negligible radial displacements of Saturn, Uranus and Neptune, which were accompanied by stirring of the eccentricities of their orbits, as a result of their initial near-resonant configuration. These chaotic effects resulted in Saturn migrating close to the 1:2 MMR with Jupiter. In such scenarios (as has been shown, e.g. Tsiganis et al. 2005), the outer Solar system may become highly chaotic. The most typical outcome of these runs was the ejection of a planet from the solar system (18 out of 25 runs), or a planet-planet collision involving one or both the icy giant planets (Uranus – Neptune) (4 out of 25 runs)[4]. In the remaining three cases, all four giant planets remained within the Solar system after

---
[4] In systems that experienced planet ejections, Uranus and Neptune were the sole ejectee in twelve and four runs, respectively. Both planets were ejected in the other two such runs. In systems that experienced planet-planet collisions,

10 Myr, albeit on significantly excited orbits. These 25 runs all involve systems where the planets started in compact configurations, with the majority having $a_{N0}$ of either 15 or 16 AU. The compact nature of these systems clearly enhances the likelihood of such chaotic/catastrophic outcomes.

At the end of the integrations, no Neptune Trojans were found in those chaotic systems in which Neptune itself survived until the end of the integrations. Upon examination of the evolution of the Trojan population in these cases, however, it was apparent that the rapid decay of that population had set in prior to Jupiter and Saturn becoming mutually resonant (or suffering strong interactions), a behaviour very similar to that observed in stable resonant systems (discussed in Section 3.1). Therefore, at least in the case of these resonant scenarios, it seems that such chaotic events are not necessary in order to have an almost complete loss of pre-formed Neptune Trojans.

### 3.3 THE CAPTURE OF FORMER NEPTUNE TROJANS AS TROJANS OF THE OTHER GIANT PLANETS

As shown by Horner & Evans (2006), once an object is moving on a dynamically unstable orbit in the outer Solar system, it is possible that they can be captured as Trojans by one or other of the giant planets. It therefore seemed interesting to examine the fate of the objects that left Neptune's control in our simulations, to see whether any were captured by the other planets.

After performing a thorough analysis of all integrations carried out, a small number of objects were found to have been captured by Jupiter, Saturn or Uranus by the completion of the integrations (Table 2). By computing the number of objects that suffered close encounters with these giant planets, we estimate that the capture likelihood of former Neptune Trojans by the other giant planets is roughly $< 5\sim10\times10^{-6}$ (Jupiter Trojans), $< \sim10^{-5}$ (Saturn Trojans), and $10^{-4}\sim10^{-3}$ (Uranus Trojans). The obtained values above are similar to those found for the capture probability of primordial trans-Neptunian objects into the Trojan clouds of those planets. We refer the reader to (Lykawka & Horner 2010) for more details on the computation of such estimations.

| Architecture | Jupiter Trojans | Saturn Trojans | Uranus Trojans |
|---|---|---|---|
| Non-MMR | 0 | 1 | 17 |
| 1:2 MMR | 0 | 0 | 12 |
| 2:3 MMR | 1 | 0 | 27 |
| 3:4 MMR | 3 | 0 | 17 |

**Table 2:** Details of objects that were identified as captured Trojans of Jupiter, Saturn and Uranus at the end of 10 Myr integrations studying the evolution of Neptunian Trojans in planetary systems with various planetary architectures, set by the initial locations of Uranus and Neptune ($a_{N0}$). The scenarios studied involved those two planets being placed in non-resonant (Non-MMR) and three distinct resonant configurations (1:2, 2:3, and 3:4 MMRs). The captured Trojans detailed above were originally members of the Neptunian Trojan clouds at the beginning of the simulations.

By employing the special algorithm described in Section 3, we were able to determine the approximate number of captured Trojans controlled by Uranus over the course of the integrations. Overall, we estimate that between 1 and 3% of former Neptune Trojans became Uranian Trojans at some point during their evolution, though few, if any, remained as such for a protracted period of time. It is likely that, had more particles been used in our simulations (and therefore more objects been available for capture), the population of captured Saturnian and Jovian Trojans would have been larger at the end of the integrations (10 Myr).

---

Uranus-Neptune encounters accounted for two cases, with the remainder consisting of one Saturn-Uranus collision and one between Jupiter and Uranus.

# 4 DISCUSSION

According to our results, a significant fraction of pre-formed Neptune Trojans would have been lost within the first 10 Myr after the formation of the giant planets, prior to any significant migration. It is important to note that this loss is greatly enhanced (sometimes as high as 100%) in scenarios where the planets form in compact systems with their orbits close to their mutual MMRs (such as the Uranus-Neptune 2:3 and 3:4 resonances studied in this work). If the pre-formed Neptune Trojans were indeed decimated in the very early stages of the Solar system, the currently observed population must represent the survivors of a (presumably much larger) populations of bodies that were captured into the planet's Trojan clouds during its outward migration. This conclusion is supported by the fact that those objects which survived for the full 10 Myr duration of our integrations did so on orbits that displayed only very modest amounts of eccentricity excitation, and little, if any, change in their orbital inclination from the initial values. In contrast, the observed Neptune Trojan population in our own Solar system contains objects spread over a wide range of orbital inclinations (ranging up to 28° as of July 2009). Our earlier work (Paper I) provides further evidence for this conclusion, as the migration itself is unable to excite the majority of pre-formed Trojans to such inclinations[5]. In addition, as can be seen by the choice of initial conditions for our pre-formed Trojans, we remind the reader that such objects could be expected to acquire only very small eccentricities and inclinations by the end of their accretion process (Chiang & Lithwick 2005), and it seems highly unlikely that the primordial population could have formed with an inclination distribution matching that seen in the observed population. Taken in concert, these results argue strongly that the observed Neptune Trojans must represent primarily a captured, rather than pre-formed, population of objects.

It is well known that high-$i$ Neptune Trojans represent at least half the population in the observational data (Sheppard & Trujillo 2006). Our results suggest that the strong instability generated by the overlapping of MMRs in compact systems could play an important role in removing the potentially vast population of pre-formed low-$i$ Neptune Trojans that presumably formed along with the planet. Coupled with the loss of such objects during the migration of the planet, this might well explain why the observed population does not display an excess at very low inclinations.

Additional support for such compact planetary systems comes from our previous results, where we showed that the capture of Trojans during the migration of Neptune over a large distance (from 18.1 AU to its current location) yielded stable Trojan populations whose properties provide a noticeably better match to those of the currently known Trojans (Paper I; Lykawka et al. 2009).

In addition, as detailed in Section 3, the smaller survival fractions and quicker lifetime decays of Neptune Trojans for systems where Uranus and Neptune and/or Saturn and Uranus were close to a mutual MMR suggest that the latter two planets can play a significant role in reducing the stability of Neptunian Trojans on short time scales. This result reinforces the idea that Uranus and Saturn played a pivotal role in shaping the stable/unstable regions of the Neptune Trojan clouds (Kortenkamp, Malhotra & Michtchenko 2004; Zhou et al. 2009; Horner & Lykawka 2010). It also implies that one can expect even more severe instabilities to afflict Neptune's Trojan clouds if all of the other giant planets were located near their mutual MMRs by the end of planet formation - a situation which a number of recent models suggest was likely the case (Thommes 2005; Papaloizou & Szuszkiewicz 2005; Morbidelli et al. 2007; Thommes et al. 2008). Indeed, our results for the most compact resonant systems (Architecture D with $a_{N0}$ = 15 AU, see Fig. 2), combined with those on the influence of system compactness on Trojan stability strongly suggest that the entire

---

[5] There is evidence that the Neptune Trojans suffer negligible inclination excitation of over timescales stretching to billions of years, once planetary migration has come to a conclusion (Brasser et al. 2004; Horner & Lykawka 2010; Lykawka et al., in preparation).

primordial local Neptunian Trojan population would be removed in only a few Myr in such systems[6].

The significant loss of Neptune Trojans on short dynamical lifetimes recorded in this work also intriguingly suggest that the formation of these objects could have been significantly retarded due to a dynamically induced shortage of building blocks around Neptune's orbit. This might in turn imply that, rather than our results suggesting a rapid decay of a pre-formed Trojan population, such a population never had the opportunity to form - leading to a greatly diminished initial population which proceeded to decay to nothing well before the planets migrated significantly.

Either way, it seems clear that violent dynamical interactions between the outer planets (as such as Jupiter and Saturn crossing their mutual 1:2 or 3:5 MMRs (Morbidelli et al. 2007 and references therein)) are not required in order that a dramatic loss of dynamically cold native Neptune Trojans occur. We have shown conclusively that dynamically stable systems are capable of causing sufficient depletion of any primordial Neptunian Trojans that few, if any, would survive long enough to experience the effects of that planet's migration.

## 5 CONCLUSIONS AND FUTURE WORK

The recently discovered population of Neptune Trojans display a wide range of inclinations, in striking contrast to predictions of their distribution made prior to their discovery. In this work, we attempt to determine whether pre-formed Neptune Trojans could have survived until the epoch of planetary migration, and if so, whether dynamical effects in the pre-migration early Solar system would cause excitation sufficient to explain the currently observed excess of high-inclination objects. We therefore study various plausible pre-migration planetary architectures with varying degrees of compactness, and examine scenarios in which Uranus and Neptune lie on both non-resonant and resonant orbits (considering the mutual 1:2, 2:3 and 3:4 MMRs).

In our simulations of non-resonant planetary systems, we observed that the dynamical erosion of our pre-formed Neptune Trojans become significant if Neptune formed closer to the Sun than ~21 AU. In these configurations, more than ten percent (up to 55%) of the pre-formed Trojans were lost over the 10 Myr of our integrations. In the 1:2 MMR resonant scenario, the general behaviour was the same - the more compact the system, the more readily Trojans were lost. However, there was a notable exception in the case when Uranus and Saturn were located close to a mutual resonance. These additional resonant perturbations led to a very severe depletion of the Neptune Trojans by the end of the integrations.

In the other resonant scenarios tested, 2:3 and 3:4 MMR, the Neptune Trojans were rapidly and severely depleted - particularly in the most compact planetary architectures. In several cases, the depletion was complete - with none of the original 20000 Trojans surviving the 10 Myr integrations. The effects of mutual resonances between Saturn and Uranus were again somewhat apparent for these scenarios, although they were significantly harder to resolve given the already severe levels of Trojan depletion.

From all of the above, it is clear that the great majority of plausible pre-migration architectures of the outer Solar system lead to severe levels of depletion of the Neptunian Trojan clouds. Coupled with the results from our earlier migration work (Paper I), this suggests that the great majority of pre-formed Trojans would have been lost prior to Neptune attaining its current location. This argues that the great bulk of the current Neptune Trojan population was likely captured during that planet's migration. Observations of the current Neptune Trojans appear to add weight to this argument, since it appears that only a population of captured Trojans can explain the wide range of orbital inclinations observed.

---

[6] Therefore, at a time well before the dynamical instabilities predicted to occur at ~600-700 Myr in the 'Nice model' (Morbidelli et al. 2007 and references therein).

In future work, it would be interesting to study the orbital evolution of pre-formed Trojans for each giant planet as a function of the initial architecture of the Solar system, and on a variety of timescales. Scenarios in which the four giant planets are located in mutual MMRs would be of particular interest. Such studies, incorporating the latest models of giant planet formation and evolution, will allow the use of the Trojan populations as a tool to examine the conditions in the early Solar system, and may well help provide an extra test of new, more detailed models. This is particularly important for the Jovian Trojan population, which provides the best observational constraints for the development and further improvement of these models.

**ACKNOWLEDGEMENTS**
We would like to thank an anonymous referee for a number of helpful comments and suggestions, which allowed us to improve this work. PSL and JAH gratefully acknowledge financial support awarded by the Daiwa Anglo-Japanese Foundation and the Sasakawa Foundation, which proved vital in arranging an extended research visit by JAH to Kobe University. PSL appreciates the support of the COE program and the JSPS Fellowship, while JAH appreciates the support of STFC.

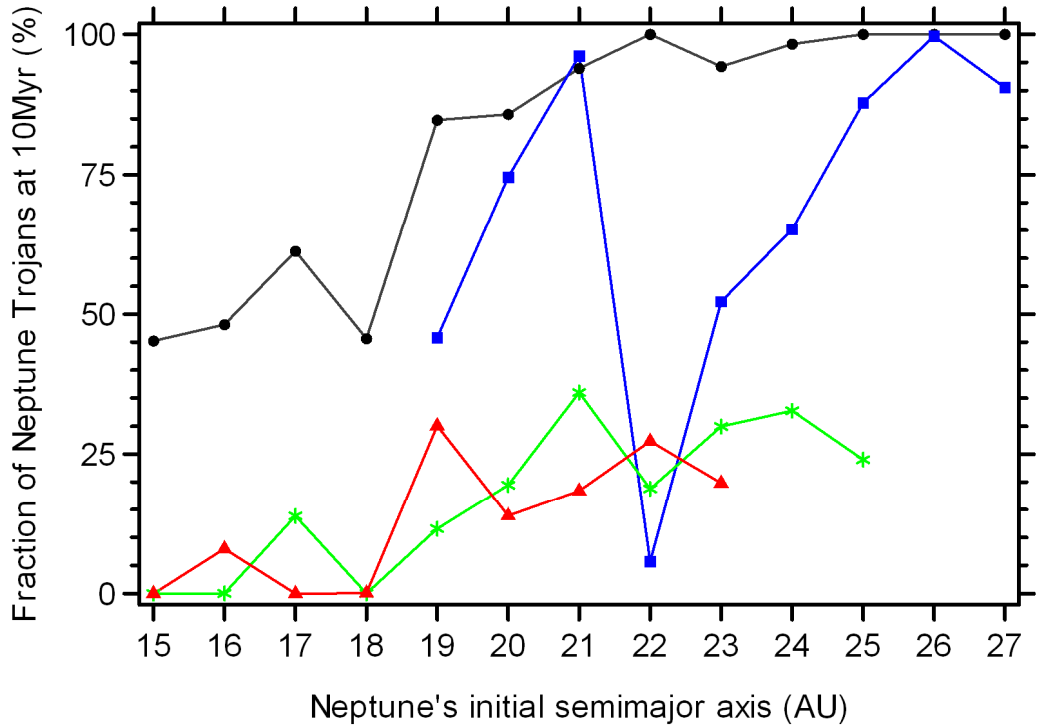

**Figure 1:** The fraction of hypothetical Neptune Trojans that survived as Trojans to the end of simulations performed for 10 Myr as a function of the initial heliocentric distance of Neptune. Each simulation followed the evolution of 2000 massless particles placed on orbits librating around Neptune's L4 and L5 Lagrange points (1000 bodies each) in a planetary system with the four giant planets (Jupiter, Saturn, Uranus and Neptune) located at heliocentric distances compatible with their pre-migration locations in current theories of Solar system formation. Each data point gives the averaged value obtained across 10 separate runs (each following 2000 particles) which differ only in the initial location of the various planets on their orbit (the orbits themselves were kept constant across each suite of 10 runs). The results from four different scenarios are shown: integrations in which the initial orbits of Uranus and Neptune were not resonant with one another are shown in black, those in which the two outer planets were located close to their mutual 1:2 MMR are shown in blue, while the data plotted in green and red detail further near-resonant runs where the planets were close to their 2:3 and 3:4 MMRs, respectively.

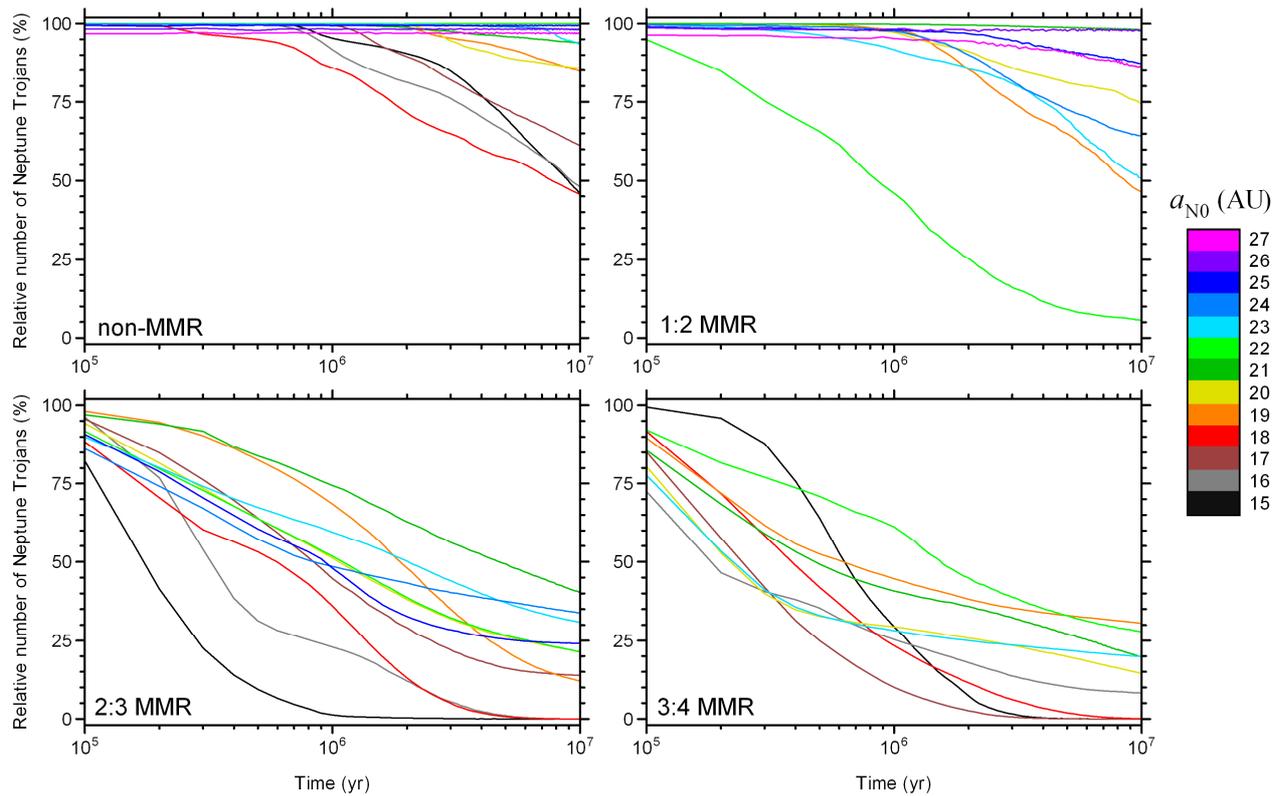

**Figure 2:** The decay of the hypothetical Neptune Trojan population as a function of time, for a variety of planetary architectures. The various lines show the evolution of the Trojans, with their colour revealing the initial semimajor axis at which Neptune was placed in those systems. The simulations employed 1000 massless bodies initially located on orbits librating around each of the Neptunian L4 and L5 Lagrange points, and were repeated ten times for each scenario using the same initial planetary orbital parameters, with the planets placed at random locations on their orbits. Each line reveals the overall result, averaged over the ten separate trials. The four panels show the results for the four different types of planetary architecture considered in this work. The top left panel reveals the results when the giant planets are located far from their mutual mean-motion resonances (top left-hand panel). The other three panels show the results when Uranus and Neptune lie close to their mutual 1:2 MMR (top right-hand panel), 2:3 MMR (lower left-hand panel) and 3:4 MMR (lower right-hand panel), respectively.